\documentclass[12pt,preprint]{aastex}

\usepackage{natbib,graphicx}

\slugcomment{Accepted to Astrophysical Journal.}

\shorttitle{HD 80606b}
\shortauthors{Langton \& Laughlin}

\begin{document}

\title{Hydrodynamic Simulations of Unevenly Irradiated Jovian Planets}

\author{Jonathan Langton}
\affil{Department of Physics\\ University of California at Santa Cruz\\
    Santa Cruz, CA 95064}
\email{jlangton@ucsc.edu}

\and
\author{Gregory Laughlin}
\affil{UCO/Lick Observatory\\ 
Department of Astronomy and Astrophysics\\
University of California at Santa Cruz\\
Santa Cruz, CA 95064}
\email{laughlin@ucolick.org}

\begin{abstract}
We employ a two-dimensional grid-based hydrodynamic model to simulate upper atmospheric dynamics on extrasolar giant planets.  The hydrodynamic equations of motion are integrated on a rotating, irradiated sphere using a pseudospectral algorithm.  We use a two-frequency two-stream approximation of radiative transfer to model the temperature forcing.  This model is well-suited to simulate the dynamics of the atmospheres of planets with high orbital eccentricity that are subject to widely-varying irradiation conditions.  We identify six such planets, with eccentricities between $e=0.28$ and $e=0.93$ and semimajor axes ranging from $a=0.0508$ A.U. to $a=0.432$ A.U., as particularly interesting objects for study.   For each of these planets, we determine the temperature profile and resulting infrared light curves in the 8-$\mu$m Spitzer bands.   Especially notable are the results for HD\, 80606b, which has the largest eccentricity ($e=0.9321$) of any known planet, and HAT-P-2b, which transits its parent star, so that its physical properties are well-constrained.  Despite the variety of orbital parameters, the atmospheric dynamics of these eccentric planets display a number of interesting common properties.  In all cases, the atmospheric response is primarily driven by the intense irradiation at periastron.  The resulting expansion of heated air produces high-velocity turbulent flow, including long-lived circumpolar vortices.  Additionally, a superrotating acoustic front develops on some planets; the strength of this disturbance depends on both the eccentricity and the temperature gradient resulting from uneven heating.  The specifics of the resulting infrared light curves depend strongly on the orbital geometry.  We show, however, that the variations on HD 80606 b and HAT-P-2b should be readily detectable at  4.5 and 8 $\mu$m using the Spitzer Space Telescope.  Indeed, these two objects present the most attractive observational targets of all known high-$e$ exoplanets.
\end{abstract}


\section{Introduction}

In July of 1994, the astronomical community was captivated by the collision of comet Shoemaker-Levy 9 and Jupiter.  The sudden deposition of energy into the Jovian atmosphere caused a dramatic response on a planetary scale, providing remarkable opportunities for observation and simulation \citep{har01, dem01} and fostering the development of a deeper understanding of the structure and behavior of gas-giant atmospheres.   Within our own solar system, such an occurrence is a one-time event.  However, the discovery of extrasolar giant planets on orbits with substantial eccentricity affords the opportunity to study atmospheric dynamics on worlds where the sudden influx of a massive dose of energy occurs on a regular basis.  With the advent of Spitzer, direct observation of such planets has become feasible.

The Spitzer telescope has been an unexpectedly useful tool in the
study of hot Jupiters, and has already been used to make a number of landmark
observations of extrasolar planets. These began with the detections of the secondary
eclipses of HD 209458 b \citep{dem05} and TrES-1 \citep{char05},
which
confirmed that hot Jupiters have low albedos and day-side effective
temperatures of order $T_{\rm eff}\sim1100\,{\rm K}$. The TrES-1 observations were
carried out at 4.5$\mu{\rm m}$ and 8$\mu{\rm m}$, and thus provided an estimate of that
planet's mid-infrared spectral slope.
High S/N observations at 16 $\mu{\rm m}$ of
the secondary transit of HD 189733 \citep{dem06} produce a similar
($1117\pm42\,{\rm K}$) day-side temperature.

More recently,
Harrington et al. (2006) have detected strong orbital phase-dependent brightness
variations
for $\upsilon$ And b at 24 $\mu{\rm m}$, which indicate a strong day-night temperature
contrast in the far IR, and suggest that heat transfer
between the hemispheres is inefficient at the level in the atmosphere that generates
the observed $24\mu{\rm m}$ emission.
Conversely, Cowan et al. (2007) found no significant orbital
phase variations at 8 $\mu{\rm m}$ for 51\,Peg b and HD\,179949 b, indicating that
for those two planets, surface flows {\it are} effective at redistributing heat at
the level probed by the 8 $\mu{\rm m}$ photosphere.
Even more recently, Knutson et al. 2007 have presented results from a continuous
30-hour Spitzer observational campaign on HD 189733. The 
resulting 8$\mu$m photometric time-series contains extraordinary
(part in $10^{-4}$) signal-to-noise and clearly shows phase-dependant
flux variations that indicate a day-night temperature contrast of order 200 K.
Of particular interest in the Kuntson et al. (2007) time series is a $F/<F> \sim 0.0005$
increase in planetary flux of $\sim 5$hr duration just past the primary transit. This
feature can be interpreted as arising from a hotspot near the dawn terminator
of the planet. 

In addition to providing infrared photometric time-series data,
the Spitzer Space Telescope can also be used to obtain mid-infrared spectra
of extrasolar planets. This has been done for the transiting planets 
HD\,209458b (Richardson et al. 2007) and HD\,189733b (Grillmair et al. 2007).
In both cases, spectra in the 7-14$\mu$m range
were obtained by subtracting the stellar spectrum measured
during secondary eclipse from the combined star+planet spectrum obtained
just outside eclipse. Model atmospheres of extrasolar planets under strong
insolation conditions (e.g. Sudarsky, Burrows \& Hubeny 2003, Marley
et al. 2006) predict that ${\rm H_2O}$ absorption features would be
prominent within the observed spectral range. Remarkably, however,
neither planet displayed the predicted signature. This null result can be
interpreted in several ways. First, it is possible that the atmospheres
are dry, perhaps because available oxygen is sequestered in CO. This
interpretation, however would require both planets to have unexpectedly
large C to O ratios. Alternately,
the effectively featureless spectrum could arise from a very clear 
atmosphere with an isothermal P-T profile down to large optical depth \citep{for06}. This
interpretation is favored by the fact that the hot Jupiters have extremely
low albedos (Rowe et al. 2006). As a third possibility, \citet{har07} suggest that
HD\,209458b may be displaying line emission at 10$\mu$m. They
interpret these features as possibly arising from silicate clouds. Any
such clouds, however, would need to be highly non-reflective in order
to be consistent with the low optical albedo.

At present, the existing observational results present an incomplete and somewhat contradictory overall
picture. It is not understood how the wind vectors and temperature distributions
on the observed planets behave as a function of pressure depth, and planetary longitude
and latitude.
Most importantly, the effective radiative time constant in the atmospheres of
short-period planets
remains
unmeasured, and as a result, dynamical calculations of the expected planet-wide
flow patterns (Cho et al. 2003, Cooper \& Showman 2005, Burkert et al. 2005,
Fortney et al. 2006, Langton \& Laughlin 2007, Dobbs-Dixon \& Lin 2007) have come to no consensus
regarding how the surface flow should appear. The lack of agreement between
the models stems in large part from the paucity of unambiguous measurements of the
basic thermal structure of the atmospheres of short-period planets.
Essentially, the problem comes down to the fact that the planets that have been observed so
far have had their spin periods tidally synchronized to their orbits, so
that each spot on the substellar surface sees a constant illumination from the parent star.
The resulting flow pattern will therefore likely be in steady state,
and the temperature distribution across the planetary surface will depend on both
the radiative timescale as well as the windspeeds at optical depth unity.

In this paper, we discuss our modeling program for determining the
hydrodynamic flows on the surfaces of planets traveling on significantly eccentric orbits.  Such planets are not spin-synchronous. Tidal effects cause the rotation rate to approximate the orbital angular velocity at periastron; the rotation is therefore pseudo-synchronous and the spin period is a sensitive function of the eccentricity \citep{hut81}.  In contrast to the typical spin-synchronous hot Jupiter, on which one hemisphere is perpetually illuminated while the other receives no insolation, the entire surface of an eccentric planet receives radiation.  Pseudo-synchronization does, however ensure that a single hemisphere receives the preponderance of the illumination during the periastron passage.  Additionally, the intensity of the radiation received varies strongly over the course of an orbit.  The insolation of planets on eccentric orbits therefore exhibits significant time variation.

To our knowledge, other modeling efforts to date have focused exclusively on
exoplanets that are in tidally synchronized orbits with a permanent day-night side 
asymmetry and essentially steady state irradiation conditions. With such bodies, it is difficult to disentangle the contributing
effects of the effective radiative time constant in the planetary atmosphere, and
the hydrodynamic flow velocities on the surface. In this paper, we propose that these
ambiguities can be resolved by observing planets that have eccentric orbits, and
which hence show different hemispheres (or partial hemispheres) to the star, and
which experience a highly variable total insolation.
We identify the highly eccentric planet HD 80606 b and the transiting planet HAT-P-2 b \citep{bak07} as being a prime targets for such observations.  We
propose that by monitoring these planets during their periastron
passages, we can improve our overall understanding of the atmospheric
properties of hot Jupiters. The planet-wide response
to the rapid increase in stellar radiation will be readily observable
in the 4.5$\mu{\rm m}$ and 8$\mu{\rm m}$ bands, and will
provide a direct simultaneous measurement of the radiative time
constant at two separate levels in the atmosphere.

The plan for this paper is as follows. In \S 2, we describe our numerical model as applied to extrasolar planets. In \S 3, we discuss the characteristics of the six planets considered in this paper, as well as relevant stellar properties of their parent stars.  In \S 4, we present the results of this model as applied to these planets.   In \S 5, we discuss the observational consequences of these results, in particular concentrating on the infrared light curves in the 8 micron Spitzer band.  We show that in general, the periastron passage produces a distinct flux peak, and that the shape of the overall curve can be used to constrain planetary characteristics such as rotation rate and atmospheric radiative time constant.  We conclude in \S 6.

\section{Numerical Model}

We begin with the primitive hydrodynamic equations of motion in two dimensions, 
assuming hydrostatic equilibrium:
\begin{eqnarray}
\frac{\partial \rho}{\partial t} &=& - \mathbf{v} \cdot \nabla \rho - \rho \nabla \cdot \mathbf{v}\\
\frac{\partial \mathbf{v}}{\partial t} &=& - \mathbf{v} \cdot \nabla \mathbf{v} - \frac{1}{\rho}\nabla p\\
\frac{\partial}{\partial t}\left(\frac{e}{\rho}\right) &=& - \mathbf{v} \cdot \nabla \left(\frac{e}{\rho}\right) - \frac{p}{\rho} \nabla \cdot \mathbf{v}. \label{energyeq}
\end{eqnarray}
We use the equation of state for an ideal gas: 
$p = (\gamma - 1)e = \rho R T$, where $R$ is the \emph{specific} 
gas constant and the adiabatic constant $\gamma = 1.4$ for a diatomic gas.  (We actually use a value of $\gamma = 1.389$, appropriate for molecular hydrogen at 900 K.)  Setting the specific heat capacity to $C_V = R/(\gamma-1)$, we can substitute

\begin{equation}
T = \frac{1}{C_V} \frac{e}{\rho}
\end{equation}
into equation (\ref{energyeq}) to obtain the hydrodynamic equations in 
terms of the dynamical variables $T, \mathbf{v},$ and $\rho$:
\begin{eqnarray}
\frac{\partial T}{\partial t} &=& - \mathbf{v} \cdot \nabla T - (\gamma-1) T \nabla \cdot \mathbf{v} \label{Teq0}\\
\frac{\partial \mathbf{v}}{\partial t} &=& - \mathbf{v} \cdot \nabla \mathbf{v} - \frac{RT}{\rho}\nabla \rho - R \nabla T \label{veq0}\\
\frac{\partial \rho}{\partial t} &=& - \mathbf{v} \cdot \nabla \rho - \rho \nabla \cdot \mathbf{v} \label{rhoeq0}.
\end{eqnarray}
In order to obtain a model appropriate to planetary atmospheric flows, we add a thermal forcing term to (\ref{Teq0}) and a Coriolis term to (\ref{veq0}), obtaining the equations of hydrodynamic motion on an irradiated, rotating sphere:
\begin{eqnarray}
\frac{\partial T}{\partial t} &=& - \mathbf{v} \cdot \nabla T - (\gamma-1) T \nabla \cdot \mathbf{v} +f_{rad}\label{Teq}\\
\frac{\partial \mathbf{v}}{\partial t} &=& - \mathbf{v} \cdot \nabla \mathbf{v} - \frac{RT}{\rho}\nabla \rho - R \nabla T -2\Omega_{rot} \sin \theta (\hat{n} \times \mathbf{v})\label{veq}\\
\frac{\partial \rho}{\partial t} &=& - \mathbf{v} \cdot \nabla \rho - \rho \nabla \cdot \mathbf{v} \label{rhoeq},
\end{eqnarray}
where $f_{rad}$ gives the thermal forcing, $\Omega_{rot}$ is the angular frequency of the planet's rotation,  $\theta$ is the latitude, and $\hat{n}$ is a unit vector normal to the planet's surface.

We calculate $f_{rad}$ using a one-layer, two-frequency radiative transfer scheme.  
For hot Jupiters, the bulk of the energy input occurs near the peak of
the stellar spectrum in the visible $\lambda_{\rm max} \sim 2.73/T_{\star} \sim 0.5 \mu$.
Energy balance, on the other hand, is maintained via thermal
re-radiation in the infrared $\lambda_{\rm max} \sim 2.73/T_{\rm pl} \sim 3 \mu$.
The separation of the corresponding blackbody distributions motivates
our use of a two-frequency scheme.
We take the mean opacity for incident stellar radiation to be $k_1$, while 
the mean opacity for outgoing long-wave radiation is $k_2$.  In this greatly 
simplified model, a thin layer of the atmosphere with thickness $dz$ and 
at a pressure depth $p$ will absorb a flux equal to

\begin{equation}
dF = \left(\frac{k_1 F_*(p)}{\cos \alpha} + k_2 \sigma T_n^4 - k_2 \sigma T^4 \right) \rho \phantom{i} dz,
\end{equation}
where $F_*(p)$ is the incident stellar flux at a pressure depth $p$, $\alpha$ is the zenith angle of the star, $T_n$ is the temperature of the layer being modeled in the absence of solar heating, and $\sigma = 5.67 \cdot 10^{-8}$ W m$^{-2}$ K$^{-4}$ is the Stefan-Boltzman constant.  $T_n$ is the result of the combination of the intrinsic planetary flux due to tidal heating and the heating supplied by the atmosphere at high optical depth, and is calculated using a one-dimensional two-stream approximation to radiative transfer similar to that used in \citet{iro05}.  The details of the $T_n$ calculation are discussed in greater detail in \S 3.

Energy conservation requires that $dF = \rho C_p f_{rad} dz$, so the thermal forcing must be
\begin{equation}
f_{rad} = \left(\frac{1}{C_p}\right)\left(\frac{k_1 F_*(p)}{\cos \alpha} + k_2 \sigma T_n^4 - k_2 \sigma T^4\right).
\end{equation}

We may approximate $F_*(p)$ by ignoring the variation in $k_1$ with temperature and pressure; this gives
\begin{equation}
F_*(p) = F_*(0) e^{-\frac{k_1 p}{g \cos \alpha}}.
\end{equation}
But $F_*(0)$ is simply the flux at the top of the atmosphere, which is
\begin{equation}
F_*(0) = (1-A)\left(\frac{L_*}{4 \pi a^2}\right) \cos \alpha,
\end{equation}
where $A$ is the Bond Albedo, $L_*$ is the stellar luminosity, and $a=a(t)$ is the time-dependent distance between the star and the planet.  Then the thermal forcing is
\begin{equation}
f_{rad} = \left(\frac{1}{C_p}\right)\left(k_1(1-A)\left(\frac{L_*}{4 \pi a^2}\right) e^{-\frac{k_1 p}{g \cos \alpha}}+k_2 \sigma T_n^4 - k_2 \sigma T^4\right).
\end{equation}
With a little bit of algebraic manipulation, this can be placed in a more transparent form:
\begin{equation}
f_{rad} = \beta \left(T_{eq}^4 - T^4\right),
\end{equation}
where $\beta = \sigma k_2/C_p$ and the equilibrium temperature $T_{eq}$ follows 
\begin{equation}
T_{eq}^4 = \left(\frac{k_1}{k_2}\right)T_{ss}^4 x^{\sec \alpha} + T_n^4,
\end{equation}
with $\sigma T_{ss}^4 = (1-A)L_*/(4 \pi a^2)$ and $x = \exp(-k_1 p /g)$.
The forcing that results from this scheme strongly depends on the choice of $k_1$, $k_2$, and $p$.  Motivated by many-layer radiative models of HD 209458 b \citep{iro05}, we choose $k_1 = 2 \cdot 10^{-4}$ m$^2$ kg$^{-1}$, $k_2 = 4 \cdot 10^{-4}$ m$^2$ kg$^{-1}$, and $p = 250 (g/10$ m s$^{-2})$ mbar, where $g$ is the acceleration due to gravity at the planet's surface.

While we believe this model represents an improvement over earlier work, it is nevertheless unable to account for a number of potentially important effects.  Two-dimensional hydrodynamics obviously cannot account for vertical atmospheric flows.  Since the infrared photosphere is well inside the radiative zone for the planets under consideration, it is likely that convective effects are negligible.  However, the inability of the model to account for vertical expansion under heating could cause some error.

In our radiative model, we make a number of simplifications are likely to cause discrepancy between our predictions and observations.  We ignore the variation of opacities with temperature and pressure, which is known to be significant.  We furthermore do not consider the variation in opacity with wavelength within the infrared regime; this implies that, for example, the 24-micron photosphere and the 3-micron photosphere occur at the same pressure level.  The planets in our simulations experience large temperature variations, which may lead to the formation and dispersal of different cloud species \citep{sud03, mar07}.  We do not, however, include the formation of clouds in our model.
 
Despite these shortcomings, this model represents a significant improvement over earlier simulations in a number of areas.  Shallow water models can achieve high resolution, but also have difficulty modeling certain qualitative aspects of the flow; temperature waves redistribute heat far more efficiently than is realistic.  Since our code employs fully compressible hydrodynamics, we are able to simulate heat transfer within atmospheric flows with greater realism than is possible using incompressible shallow-water dynamics.

Furthermore, our radiative scheme provides a more realistic model of the thermal forcing than the Newtonian relaxation employed in earlier hydrodynamic simulations \citep{cho03, coo05, lan07}.  While Newtonian relaxation is a reasonable approximation when the temperature is close to the equilibrium temperature, it becomes dramatically less accurate as the temperature perturbations grow larger.  In the forcing regimes expected on extrasolar planets, the Newtonian approximation can overestimate the actual rate of temperature change by as much as a factor of 3.  Our radiative model therefore allows us to treat the stellar heating more accurately than previous simulations; this is particularly important in cases where the thermal forcing varies significantly with time, either due to non-synchronous rotation or to a highly eccentric orbit.

Equations (\ref{Teq}) through (\ref{rhoeq}) are numerically integrated using the vector spherical harmonic transform method \citep{ada99}.  We use a longitude-latitude grid with T171 resolution ($512   \times 257$), sufficiently fine to resolve the details of turbulent flow.  Numerical stability is maintained using fourth-order hyperviscosity \citep{pol04}, which prevents the spurious addition of energy at small scales.  This is accomplished by adding a term of the form $B \nabla^4 \phi$ to each of the time-evolution equations, where $\phi$ is the evolving quantity.  To minimize the effect of this numerical dissipation on the flow evolution, $B$ is generally chosen to be as small as possible, while still preventing unphysical spectral ringing or numerical instability.  In general, this requires $B \sim 10^{-12} a^4,$ where $a$ is the planetary radius.  This results in a decrease in $\phi$ of $\sim$ 5 \% per time step at the smallest spectral scales; with such strongly-varying heating, anything less allows pernicious behavior.

The planetary atmosphere is initially at constant pressure, isothermal, and nearly at rest, with only a small random initial flow to break the unphysical north-south symmetry.  Since the resulting flow is often turbulent, the specific evolution of the dynamical variables changes from one run to the next.  Nevertheless, the qualitative properties of the flow are not altered by small changes to the initial wind flow.  The initial temperature is chosen so that it approximately equals the average equilibrium temperature; this prevents the initial thermal forcing from inducing the large winds that result if, for example, the initial temperature is $T_n$ \citep{coo05}.  While other groups have obtained supersonic flow, these seem to occur in simulations where either the night-side equilibrium temperature is unusually low \citep{dob07} or the planet is initially cold \citep{coo05}.  The first situation maintains an artificially high equilibrium temperature gradient; the second greatly amplifies the effect of the initial heating, particularly if the stellar heating is not introduced gradually.  In either case, there is good reason to believe that the actual wind speeds may be subsonic.  The planet is allowed to evolve from its initial state until a dynamical equilibrium is reached -- typically $\sim 100$ days -- and then the temperature, density, wind velocity, and relative vorticity are recorded.

\section{Planetary Characteristics}
Our code requires a number of planetary characteristics as input parameters.  The mass and radius of the planet must be specified in order to determine the surface gravity and therefore the appropriate pressure at the infrared photosphere.  The orbital period, semi-major axis, and eccentricity of the planet's orbit are required to calculate the time-dependent irradiation conditions.  Additionally, the longitude of periastron is needed to produce the light curves.  With the exception of the planetary radius, all of this information can be found in \citet{but06}.  The radii are determined using the planetary structure model of \citet{bod03}.  To determine correctly the mass and the light curve, it is also necessary to know the orbital inclination.  In general, however, this can be constrained only in the case of a transit.  Nevertheless, inclinations close to $90^\circ$ are more probable.  We therefore take the minimum value of the planetary masses; that is, $\sin i = 1$.  The effect of the planetary mass enters the simulation primarily by determining the magnitude of the gravitational acceleration at the planetary surface.  Test runs of our model varying gravity by up to a factor of 10 produce little change in the resulting flow patterns.  Since a two-dimensional model of a radiative atmosphere necessarily assumes hydrostatic balance and vertical stratification, the small impact of gravity is not surprising.  In any case, it is clear that in our model, the actual mass of the planet is of negligible importance; the minimum mass assumption therefore introduces no significant inaccuracies.

The planet's rotation rate is also an important parameter, necessary to determine both the substellar point and the strength of the Coriolis effect.  In the case of a close-in circular orbit, planetary rotation is expected to be synchronous, so that the orbital period and the rotation period are identical.  If the orbit is even slightly eccentric, however, spin-synchronization is broken.  Tidal effects are strongest at periastron, and therefore the preferred rotation rate is approximately the angular frequency at periastron.  The result is a state of spin pseudo-synchronization, wherein the rotation rate is a well-defined function of the eccentricity  \citep{hut81}:
\begin{equation}
\tau_{rot}=\tau_{orb}\left(\frac{(1+3e^2+\frac{3}{8}e^4)(1-e^2)^{3/2}}{1+\frac{15}{2}e^2+\frac{45}{8}e^4+\frac{5}{16}e^6}\right).
\end{equation}

Recently, Ivanov \& Papaloizou (2007) have advanced an alternate
theory for spin-synchronization
which predicts that planets in highly eccentric orbits will have spin
angular velocities that are close
to 1.55 times the circular orbit angular velocity at periastron. As
will be clear from our simulations
in this paper, long-duration observations with the Spitzer Space
telescope can in principle be used to
distinguish between the two models for spin pseudo-synchronization.
In this initial work, however,
we adopt the Hut (1981) theory for our assumed planetary spin rates.

In addition to the planet's physical and orbital characteristics, we also must determine the equilibrium night-side temperature, $T_n$ at the simulation pressure depth $p=250 (g/10$ m s$^{-2})$ mbar.  This represents a combination of the planet's intrinsic luminosity due to tidal dissipation and gravitational contraction and the redistribution of incident stellar radiation at high optical depths.  With the exception of HD 80606 b, which is characterized by an extremely high tidal luminosity and relatively low average insolation, the night side temperatures of the planets here considered are largely determined by the incident radiation.

We determine $T_n$ using a one-dimensional plane-parallel two-stream algorithm with $N=100$ vertical zones.  In keeping with the radiative model described in \S 2, we consider only two frequencies and neglect the variation in opacities with temperature and pressure.  The results of these calculations are therefore not precise, but are likely to give reasonable estimates of the atmospheric temperature in the absence of direct stellar heating.  At the upper boundary, we impose a downward flux equal to the time-averaged stellar radiation
\begin{equation}
<F_\downarrow> = \frac{L_*}{16\pi} \left<\frac{1}{r^2}\right>=\frac{L_*}{16 \pi a^2} \frac{1}{\sqrt{1-e^2}}.
\end{equation}
At the lower boundary, which we set at $p=5.0 (g/10$m s$^{-2}$) bar, we require a net upward flux consistent with the planet's intrinsic luminosity.  We then allow the atmosphere to come to radiative equilibrium, thereby determining the temperature at the base of the atmosphere.  We then maintain the lower boundary at this temperature, and recalculate the equilibrium temperature at each layer in the absence of stellar radiation.  This quantifies the response of the upper atmosphere to heating by high-temperature interior layers.  We take $T_n$ to be the atmospheric temperature at the layer corresponding to the pressure depth used in our simulations.  The planetary parameters, including $T_n$ and $\tau_{rot}$, are summarized in Table \ref{planettab}.

Finally, the characteristics of the parent star are necessary to determine the magnitude of the incident stellar radiation, as well as the relative planet-star infrared flux as a function of wavelength.  From \citet{but06}, we obtain the stellar mass, radius, temperature, and luminosity, although our model only requires the temperature and luminosity.  These data are compiled in Table \ref{startab}.

We wish to focus particular attention on two planets: HD 80606 b and HAT-P-2 b.  With $e>0.9$, HD 80606 b is the most eccentric planet known by a significant margin.  Because both the rotation rate and the characteristics of the periastron passage depend strongly on the eccentricity, it is especially important to obtain an accurate value.  Using both the 55 radial velocities published in support of the initial detection \citep{nae01} and a further 46 radial velocities published by \citet{but06}, we obtain best-fit values for the relevant planetary characteristics:  the orbital period $\tau_{orb}=111.4298
\pm 0.0032 {\rm d}$, the eccentricity $e=0.9321 \pm 0.0023$, and the  longitude
of periastron $\varpi=300.264^{\circ}\pm0.35^{\circ}$.

HD 80606's  mass is estimated to lie between 0.9 and 1.1 $M_{\odot}$ (Naef et al. 2001, Butler et al. 2006). Adopting
$M=1M_{\odot}$, we derive $M_{\rm pl}\sin(i)=4.18 \,M_{\rm Jup}$. Our planetary
models (Laughlin et al. 2005) predict a radius of $1.10\pm0.02\,R_{\rm Jup}$. The
uncertainty in $R_{\rm pl}$ is small
because the planet lies in the mass regime where the radius is an
extremely weak function of the mass. Furthermore, the planet is sufficiently 
massive enough that while it experiences considerable internal
heating associated with its strong tidal dissipation, the heating
(which we estimate below to be $L\sim10^{28}$ ergs) is expected to have a negligible
impact on the planet's physical radius (Gu, Bodenheimer, \& Lin 2004).
The well-determined radius is important because it largely eliminates
an unknown free parameter when we model the hydrodynamic response.

Because of the extreme eccentricity and relatively large semi-major axis of HD 80606 b's orbit, tidal heating becomes a significant contribution to the overall planetary temperature.  \Citet{wum03} posit that  HD 80606 b's large eccentricity was initially produced by resonant interaction with HD 80607, a binary companion to HD 80606 with a wide separation $\sim 1000$ A.U.  Wu and Murray show
that if HD\,80606 b initially formed in an $a_{\rm init}= 5$ A.U. orbit that was highly inclined
to the binary plane, then the Kozai resonance \citep{koz62} would
have quickly driven the planet's eccentricity to $e\sim0.993$. In the Wu-Murray theory, successive
cycles of high eccentricity and attendant tidal orbital decay acted to decrease the semi-major axis. Once $a$ dropped below $\sim3$ A.U., the Kozai resonance was destroyed by general relativistic precession. Over the last few Gyr, the planet has undergone continuous tidal dissipation, which has decreased $a$ and $e$ to their current values.  By considering the amount of orbital energy lost over this time period, \citeauthor{wum03} arrive at a tidal luminosity $L_{tid} \approx 10^{28}$ erg.  For the sake of comparison, the average incident stellar radiation absorbed by the planet is $L_{rad} = 3 \cdot 10^{27}$ ergs.  It is likely, therefore, that the dominant contribution to $T_n$ on HD 80606 b arises from eccentricity tides rather than convective redistribution of incident radiation.

HAT-P-2 b also deserves special consideration.  It was discovered via the transit method by \citet{bak07}.  Because it transits its parent star, its mass $M=8.17 M_J \pm 0.72 M_J$ and radius $R = 1.18 R_J \pm 0.16 R_J$ are well-constrained.  In particular, the $\sin i$ degeneracy is broken, since the inclination $i = 90.0^\circ \pm 1^\circ$ has been directly measured.  Furthermore, we have direct determination of the radius; this is particularly valuable given the divergence of observed radii from models in a number of cases \citep{sat05, bod03, lau05}.  HAT-P-2b orbits a bright (V=8.7) star with a period of 5.63 d; furthermore, its orbital geometry  ($\varpi = 185^\circ$), ensures that the periastron passage is midway between the primary and secondary transits and that all three events are observable within a 36-hour period.   HAT-P-2b is therefore uniquely well-suited for observation.

\section{Dynamical Evolution of the Planetary Atmospheres}
\begin{figure}
\begin{center}
\includegraphics[clip = true, bb=0in 5in 9in 16in,trim = 0in 3in 1in  0in, width=6in]{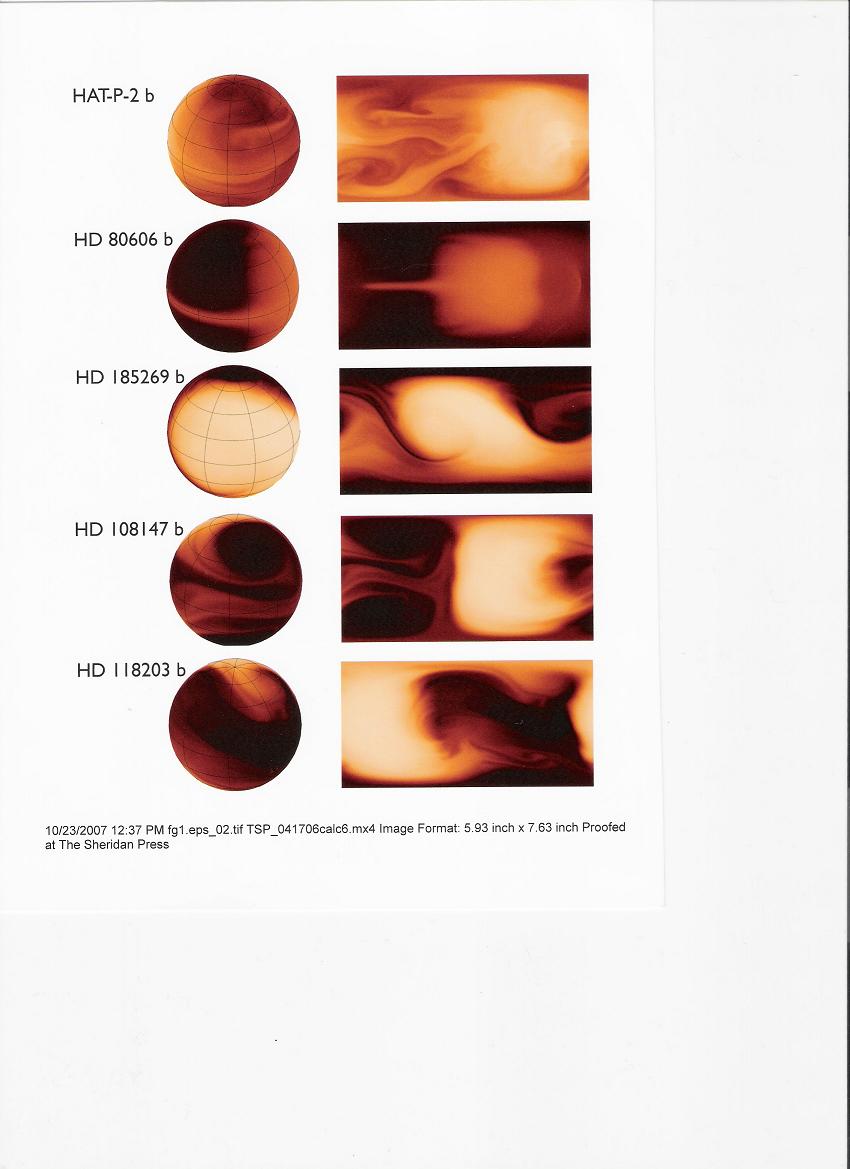}
\end{center}
\caption{Temperature distributions at representative (yet photogenic) positions in the orbit of each planet modeled.  The globes on the left show a three-dimensional representation of the temperature; the plots on the right show an equirectangular representation.  The temperature ranges in each plot are are as follows: HAT-P-2 b: 950 K to 1500 K; HD 80606 b: 710 K to 890 K; HD 185269 b: 930 K to 1530 K; HD 108147 b: 700 K to 1390 K; HD 118203 b: 970 K to 1770 K.}
\label{sixplanets}
\end{figure}

Representative atmospheric temperature distributions for each planet considered in this paper, with the exception of HD 37605 b, are presented in Figure \ref{sixplanets}.  Our simulations demonstrate that eccentric exoplanets exhibit great diversity in their atmospheric flow patterns and temperature distributions, with eccentricity, periastron distance, orbital period, and stellar luminosity all playing important roles.  Nevertheless, all of the planets here considered display significant subsolar heating at periastron, with temperature increases ranging from 300 K to 700 K.  Planets with high eccentricities and strong periastron heating may also develop high-amplitude superrotating acoustic waves in response to a high temperature gradient near the eastern terminator.  Due to the temperature dependence of the sound speed, the acoustic waves emitted inside hotter regions overtake those emitted at the cooler terminator, causing a region of very high density only a few zones wide.  This ``acoustic compression'' produces a front propogating eastward, maintaining its intensity as it is repeatedly focused by reflections off of the cooler, denser polar air.

It is not immediately obvious why our model predicts the formation of these acoustic disturbances on some planets but not on others.  The orbit of HD 37605 b is considerably more eccentric ($e=0.737$) than that of HAT-P-2 b ($e = 0.507$), yet HAT-P-2 b exhibits a clearly defined acoustic disturbance which is absent from HD 37605 b.  The periastron flux received by HD 80606 b is smaller than that received by HD 118203 b and HD 185269 b, which orbit more luminous stars.  Furthermore, HD 80606 b has a comparatively small temperature variation and correspondingly slower wind speeds.  Yet the stellar heating received by HD 80606 b generates large-amplitude acoustic waves, whereas the flow on HD 118203 b and HD 185269 is primarily vortical, with no large-scale acoustic fronts.  While we have not investigated this issue in detail, our calculation of predicted light curves suggests that the subsolar heating rate at periastron is the determinating factor in the formation of a strong acoustic front.  Sufficiently rapid heating at periastron produces a large local temperature gradient which, as described above, generates a region of high density and pressure which then propagates acoustically; slower periastron heating causes a shallower temperature gradient, less acoustic compression, and therefore any acoustic waves are much lower amplitude.

\begin{figure}
\begin{center}
\includegraphics[clip = true, bb=0in 5in 9in 16in,trim = 0in 3in 1in  0in, width=6in]{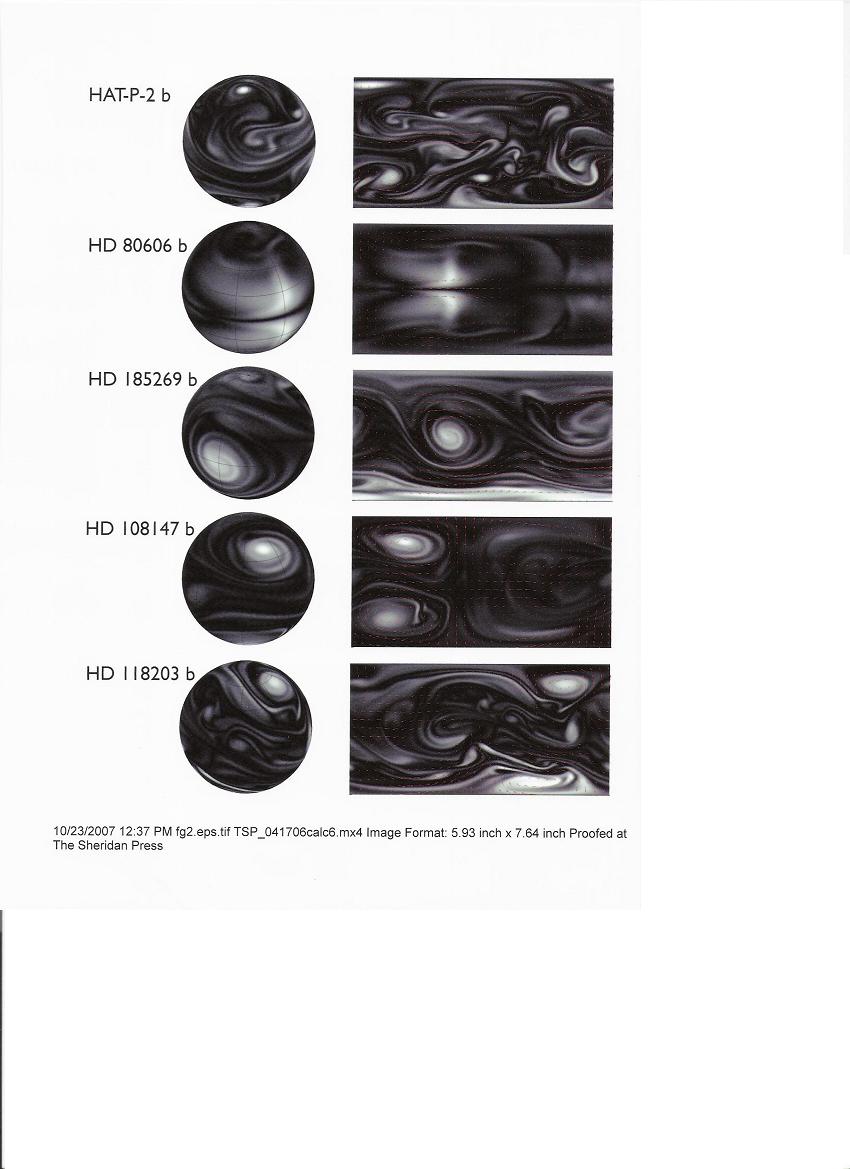}
\end{center}
\caption{Relative vorticity distributions and wind flow patterns for each planet modeled.   The globes on the left show a three-dimensional representation of the (absolute value of) relative vorticity. The plots on the right show an equirectangular plot of the wind velocity field superimposed on the vorticity distribution.   The flow pattern is taken at the same time as the temperature distribution in Figure  \ref{sixplanets}.  The maximum wind speeds in each plot are as follows: HAT-P-2 b: 1050 m s$^{-1}$; HD 80606 b: 380  m s$^{-1}$; HD 185269 b: 1410  m s$^{-1}$; HD 108147 b: 930  m s$^{-1}$; HD 118203 b: 1100  m s$^{-1}$.  The maximum relative vorticities are: HAT-P-2 b: 1.1$\cdot 10^{-4}$ s$^{-1}$; HD 80606 b: 1.8$\cdot 10^{-5}$ s$^{-1}$; HD 185269 b: 1.2$\cdot 10^{-4}$ s$^{-1}$; HD 108147 b: 1.3$\cdot 10^{-4}$ s$^{-1}$; HD 118203 b: 1.1$\cdot 10^{-4}$ s$^{-1}$.  (Note that the relative vorticity at a particular point corresponds to the angular frequency of rotation of a paddlewheel placed in the flow at that point.)}
\label{windplots}
\end{figure}

In order to elucidate the dynamics of the flow itself, we show in Figure \ref{windplots} the vorticity distribution on five of the six planets modeled; again, we omit the results for HD 37605 b. The distributions are shown at the same time and, in the case of the globe view, from the same vantage point as in Figure \ref{sixplanets}.  A quiver plot superimposed on the equirectangular map shows the direction and relative wind speed of the flow itself.

From both the temperature and vorticity distributions, it is clear that  with the exception of HD 80606 b, each planet exhibits large-scale turbulent eddies.  Given the extreme eccentricity of this planet's orbit, this is perhaps surprising.  However, the lack of turbulence can be explained by the relatively low temperature contrast (which we take to be $T_{max}-T_{min}$, where $T_{max}$ and $T_{min}$ are respectively the global maximum and minimum temperatures at a given moment) seen on HD 80606 b.  While this temperature contrast reaches 500 K momentarily, the rapid periastron passage prevents the planet from receiving enough heat to sustain these temperature differences; within 72 hours of periastron, the temperature difference has fallen to 250 K, continuing to drop after that.  In comparison, the temperature difference on HAT-P-2b exceeds 1100 K at periastron, with an average value of 540 K.  While we have not run a sufficient number of simulations to be precise, it appears that \emph{sustained} temperature differences greater than 250 to 300 K are required to generate large-scale vortical structure.  It is important that this temperature contrast be prolonged; while brief exposure to a large temperature gradient induces divergent flow, it appears that it takes some time for the Coriolis force to produce a large and persistent rotational component.

It is interesting to note here the significance of $T_n.$  For a particular amount of stellar heating, as $T_n$ increases, the induced temperature gradient becomes smaller, and both wind speed and turbulence decrease.  We find, for example, that a lower value of tidal heating on HD 80606 b, leading to $T_n = 400$ K provides a sufficiently cool background that HD 80606 b develops a temperature difference which remains over 300 K for 10 d after periastron, thereby producing turbulent flow comparable to the other planets.  Similiarly, a test run on HD 118203 b with $T_n = 1960$ K (a change with absolutely no physical justification!) reduced the maximum temperature contrast on that planet to 200 K, stifling the development of the hemisphere-scale vortices so prominent in our earlier simulation.

The intrinsic temperature of the planet then rises to a question of first-order importance.  While the infrared observations resulting from the evolution of a particular temperature distribution are not sensitive to small-scale differences in the flow, a global-scale qualitative shift from laminar to turbulent flow fundamentally alters the temperature distribution, with possibly significant observational impact.  Given the importance of tidal heating in determining the intrinsic luminosity, in addition to the unsuitability of an equilibrium model of tides to predicting the tidal dissipation on these highly dynamic objects, further development of a dynamical theory of tides seems to be a promising field for future investigation.

\section{Infrared Emission Curves}
\begin{figure}
\begin{center}
\includegraphics[clip = true, bb=0in 5in 9in 16in,trim = 0in 14in 8in  0in, width=5in]{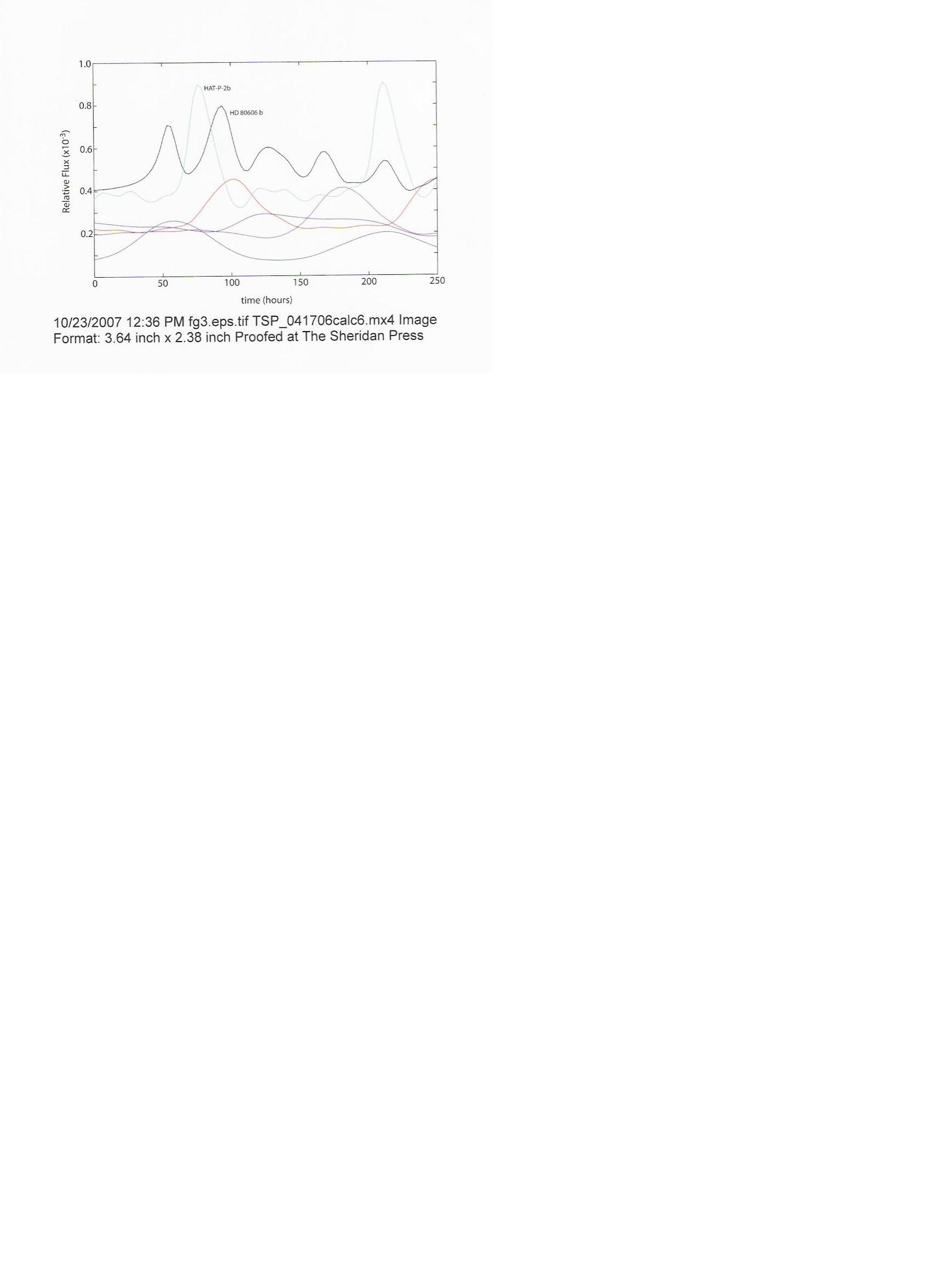}
\end{center}
\caption{8-$\mu$m infrared emissions for all six planets modeled over a 250-hour period.  The cyan curve shows the light curve for HAT-P-2 b; the black curve, HD 80606 b; the red curve, HD 118203 b; the purple curve, HD 108147; the cerulean curve, HD 185269 b; the navy blue curve, HD 37605.  These light curves were prepared under the assumption that $i=90^\circ$.}
\label{IRcurves}
\end{figure}

\begin{figure}
\begin{center}
\includegraphics[clip = true, bb=0in 5in 9in 16in,trim = 0in 4in 1.5in  0in, width=6in]{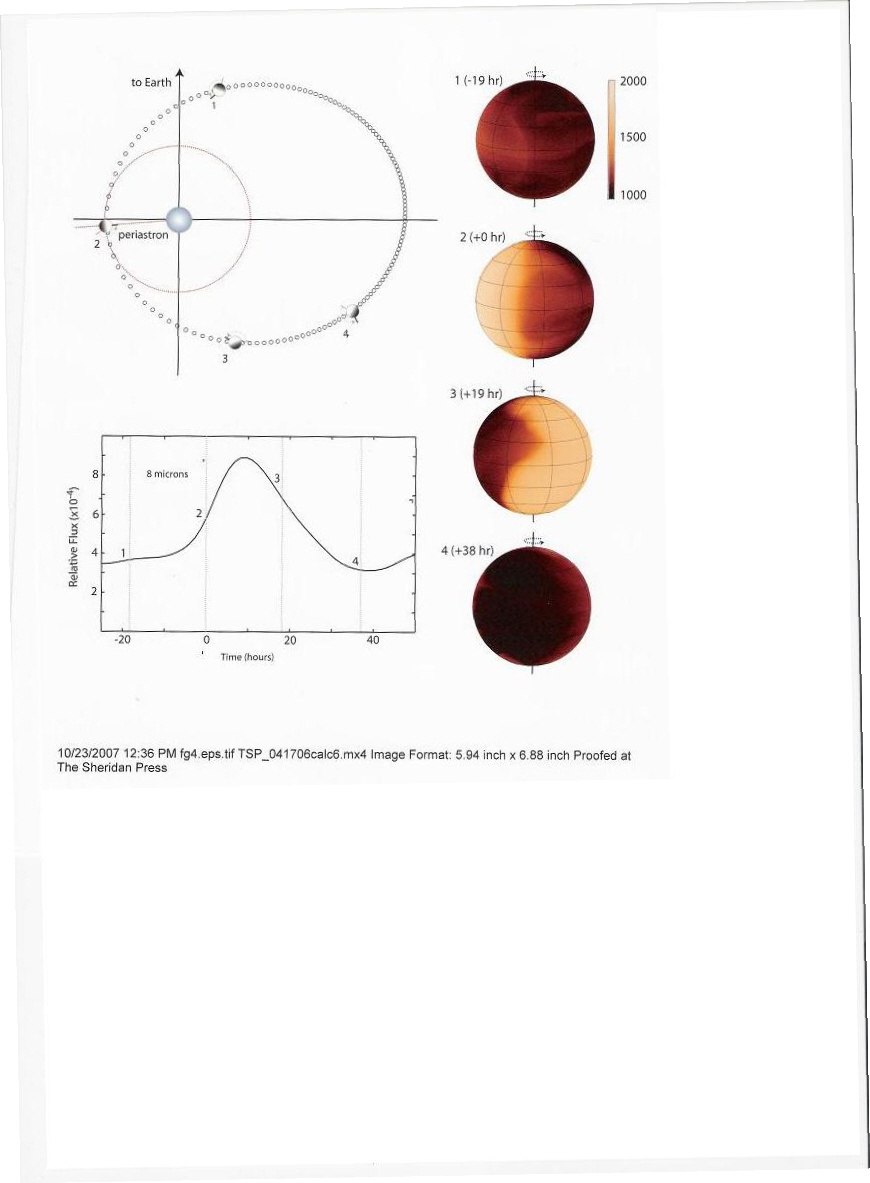}
\end{center}
\caption{A detailed representation of the results of the HAT-P-2 b simulation.  The small points on the orbital diagram show the position of the planet at one-hour intervals; the larger globes (labeled 1 through 4) show the position of the planet at times $t = -19 \rm{h}, 0 \rm{h}, +19 \rm{h}, +38 \rm{h}$ relative to periastron.  The black poles on these globes show the rotation of a stationary point on the planetary surface; the gray poles indicate the amount of rotation that has occurred over a 19-hour period.  
The hemispheric plots on the right show the atmospheric temperature distribution at the corresponding times.  The light curve shows the planet's infrared emission in the 8 $\mu$m band from 25 h before to 50 h after periastron.}
\label{HATP2b}
\end{figure}

\begin{figure}
\begin{center}
\includegraphics[clip = true, bb=0in 5in 9in 16in,trim = 0in 4in 1in  0in, width=5in]{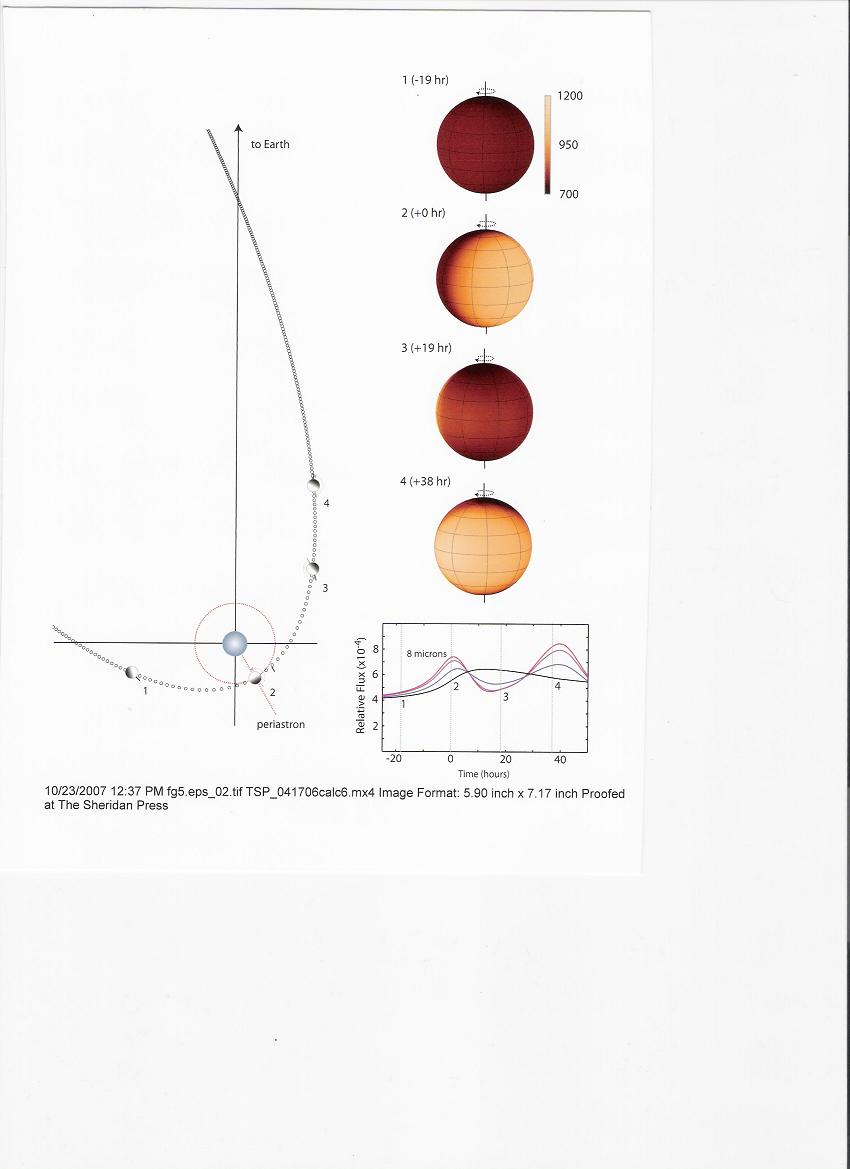}
\end{center}
\caption{A detailed representation of the results of the HD 80606 b simulation.  The small points on the orbital diagram show the position of the planet at one-hour intervals; the larger globes (labeled 1 through 4) show the position of the planet at times $t = -19 \rm{h}, 0 \rm{h}, +19 \rm{h}, +38 \rm{h}$ relative to periastron.  The black poles on these globes show the rotation of a stationary point on the planetary surface; the gray poles indicate the amount of rotation that has occurred over a 19-hour period.  
The hemispheric plots on the right show the atmospheric temperature distribution at the corresponding times.  The light curves show the planet's infrared emission in the 8 $\mu$m band from 25 h before to 50 h after periastron.  The red curve corresponds to an orbital inclination of $i = 90^\circ$ (in which the orbit is being viewed edge-on), the purple to $i=60^\circ$, the blue to $i = 30^\circ$, and the black to $i = 0^\circ$.}
\label{80606}
\end{figure}

The flux $F_\lambda$ emitted by a planet with some temperature distribution $T(\phi, \theta)$ between wavelengths
$\lambda$ and $\lambda+d\lambda$ is
\begin{equation}
F_\lambda = \int B_\lambda(T) \cos \alpha \cos \theta r_p^2 d\theta d\phi d\lambda,
\end{equation}
where $B_\lambda(T)$ is the Planck function for emission at a wavelength of $\lambda$ by
a blackbody at temperature $T$, and $\alpha$ is the angle between the surface normal
and a vector directed from the planet to the observer.  The integral is performed over the
hemisphere visible to the observer, that is, where $\cos \alpha > 0$.  In the case of a
grid-based model, this can be discretized as follows:
\begin{equation}
F_\lambda = \sum \bar{B}_\lambda(T) \cos \alpha \cos \theta r_p \Delta \theta \Delta \phi,
\label{curveeq}
\end{equation}
where
\begin{equation}
\bar{B}_\lambda(T) = \int B_\lambda(T) d\lambda.
\end{equation}
We apply equation (\ref{curveeq}) to the temperature distributions considered in \S 4 to produce planetary infrared light curves; the results are shown in Table \ref{resultstab} and Figures \ref{IRcurves}, \ref{HATP2b}, and \ref{80606}.  In Figures \ref{HATP2b} and \ref{80606}, the small points on the orbital diagram show the position of the planet at one-hour intervals; the larger globes (labeled 1 through 4) show the position of the planet at times $t = -19 \rm{h}, 0 \rm{h}, +19 \rm{h}, +38 \rm{h}$ relative to periastron.  The black poles on these globes show the rotation of a stationary point on the planetary surface; the gray poles indicate the amount of rotation that has occurred over a 19-hour period.  
The hemispheric plots in the right portion of the figures show the atmospheric temperature distribution at the corresponding times.  The light curves shows the planets' infrared emission in the 8 $\mu$m band from 25 h before to 50 h after periastron.

These light curves vary widely in their shape, range, and time scales, as is expected given the range of irradiation conditions and orbital parameters.  All show some degree of interesting structure; none can be well-fit  by the simple sinusoidal shape predicted by earlier models \citep{coo05, lan07}.  We note two particularly promising candidates for observation.  HD 80606 b, which reaches a peak flux $F/F_* = 8.0\cdot 10^{-4}$ in the 8-$\mu$m band, will be observed by the Spitzer Space Telescope during its periastron passage in November 2007.  HAT-P-2 b, which exhibits even larger flux variation over the course of its orbit, provides an equally attractive target for observation.  The flux variations for both objects are sufficiently large to be comfortably resolved by Spitzer \citep{char05, har06a, har07}.  The other four planets, HD 37605 b, HD 108147 b, HD 118203 b, and HD 185269 b, present significantly greater observational challenges.  However, it is not infeasible that the light curve variations on some of the brighter planets -- HD 118203 b and HD 108147 b -- could be resolved with sufficiently long integration times.

\section{Conclusion}

The advent of the Spitzer Space Telescope and the discovery of short-period
planets on significantly eccentric orbits has presented astronomers with 
the opportunity to observe time-dependent weather patterns on extrasolar
planets. Giant planets on eccentric orbits that experience strong variations
in insolation have the potential to tell us a great deal about the structure,
dynamics and chemistry of Jovian-like atmospheres. Eccentric planets can be
profitably compared with short period planets on spin-synchronous circular
orbits that are likely to have largely steady-state patterns of surface flow,
and which are currently being observed with Spitzer.

Furthermore, when eccentricities are extreme -- as in the case of HD 80606 b  --  we are 
afforded the ability to directly observe planets that spend the bulk of their
orbit in a region where the radiation environment is similar to that 
experienced on Earth. With current instruments, such planets would be completely
beyond the possibility of direct observation.

Eccentric planets also have the best potential to give us an understanding
of the dissipative mechanisms at work in planetary interiors. A planet such as
HAT-P-2 b should be generating a very significant luminosity as a result of the
strongly time-dependent tidal force on the planet. 
By observing the depth of the primary transit in the 
infrared, one can get a direct measurement of the planetary effective 
temperature, which can then be used to infer the amount of dissipation within
the planet, and to put interesting constraints on the tidal quality 
factor, $Q$ associated with dynamical (as opposed to equilibrium) tides.

To take a first step toward understanding what to expect from the infrared
light curves of short-period eccentric planets, we have created a
two-dimensional hydrodynamic model with radiative forcing appropriate to
the time-varying illumination received for a specified orbit. Our model
handles radiation transfer in an approximate, but hopefully qualitatively
correct manner by separating the transport of optical and IR radiation.
We apply this model to six known giant planets on eccentric orbits with
$e\gtrsim0.3$. The irradiation intensity on these planets varies by a factor
of three for the least eccentric planets up to a factor of 800 for HD 80606b.
The effect of this tremendous variability on atmospheric flows is profound;
the dominant feature in the evolution of each atmosphere studied is the
subsolar heating through the perastron passage, with the subsequent flow being
primarily driven by this dramatic forcing. For each planet, we find 
significant spatial and temporal variability in the state of the atmosphere, 
with turbulent flow being a common feature. Additionally, rapid increase
in the thermal forcing leads to superrotating acoustic shocks, 
as seen in the atmospheres of HD 80606 b and HAT-P-2b.

Among the planets that we have modeled, HAT-P-2b and HD 80606b stand as the most
promising candidates for
time-series infrared observations in Spitzer's 8-$\mu {\rm m}$
band, and among these two planets, HAT-P-2 b is the best candidate. HAT-P-2b's
secondary transit provides an opportunity for an overall flux calibration, and
its orbital geometry (with $\varpi=185^{\circ}$) places the most interesting 
features of the light curve on a relatively short 35-hour segment within 
the full 5.6-day orbit. 
During this time span, we predict that the relative infrared
flux will vary from $3.2\times10^{-4}$ to $9.0\times10^{-4}$.

HD 80606b, on the other hand, is predicted to produce a relative infrared
flux ranging from $3.8 \times 10^{-4}$ to $8.0 \times 10^{-4}$. This variation 
easily satisfies Spitzer's sensitivity limit, and for this planet, 
the observational situation is greatly
aided by HD 80607, the wide binary companion of near-identical spectral type 
to the parent star HD 80606. 
Billions of years ago, HD 80607 enabled Kozai migration.
Today, it provides a stable photometric reference to gauge the rapid
infrared flux rise from the planet during periastron passage. A measurement
of the slope
of this rise will allow a direct estimate of the atmospheric radiative
time constant at the 4.5-$\mu$ and 8-$\mu$ photospheres.

A short-period Jovian planet on an eccentric orbit likely presents one of
the Galaxy's most thrilling sights. One can imagine, for example, how
HD 80606 b appears during the interval surrounding
its hair-rising encounter with its parent star. The blast of periastron
heating drives global shock waves that reverberate several times around
the globe. From Earth's line of sight, the hours and days 
following periastron are
characterized by a gradually dimming crescent of reflected starlight, 
accompanied 
by planet-wide vortical storms that fade like swirling embers as the planet
recedes from the star. It's remarkable that we now have the ability to 
watch this scene (albeit at one-pixel and two-frequency resolution) 
from a vantage
several hundred light years away.

Indeed,
the observational and theoretical allure of short-period non-synchronous
planets is worthy of much more detailed study than we have presented here.
Our next steps on this problem, therefore, will be to (1) upgrade our
radiative transfer scheme to allow many more frequency intervals to be
modeled, and (2) incorporate the higher-resolution radiative transfer
into a full three-dimensional hydrodynamic treatment.

\acknowledgments

We are grateful to Drs. Peter Bodenheimer, Adam Burrows, Drake Deming, Dan Fabrycky, Jonathan Fortney, and Mark Marley for useful discussions.  We thank the referee for useful comments.  This research has been supported
by the NSF through CAREER Grant AST-0449986, and by the NASA Planetary Geology and Geophysics Program through Grant NNG04GK19G.  The code used for the simulations included the SPHEREPACK 3.0 subroutines provided by UCAR.

\begin{table}
\begin{footnotesize}
\begin{center}
\begin{tabular}{|r|c|c|c|c|c|c|c|c|c|c|}
\hline  &$\tau_{orb} (d)$&$\tau_{rot} (d)$& $a$ (AU)&$e$&$i$&$\varpi$  &$\frac{M_{pl} \sin i} {M_J}$  &$\frac{R_{pl}}{R_J}$&$T_{n}$ (K)\\ 
\hline  HD 80606 b	&111.449&1.72&0.432 &0.9321&--&$300.3^\circ$& 4.18  &1.1&720\\ 
\hline HD 37605 b&$54.22$&$6.76$&0.261&0.737&--& $211.6^\circ$&2.86&1.1&320\\
\hline HD 108147 b&10.90&3.50&0.102&0.53&--&76.7&0.261&0.9&670\\
\hline HAT-P-2 b&$5.63$ &$1.96$ &0.0685 &0.507 &$90^\circ$ & $185^\circ$&8.17&1.18 &1010\\
\hline HD 118203 b&6.134&3.85&0.0703&0.309&--&155.7&2.14&1.1&960\\
\hline  HD 185269 b	&6.8380&4.61&0.0766 &0.280&--&$100^\circ$& 0.91  &1.0&935\\
\hline
\end{tabular}
\end{center}
\end{footnotesize}
\caption{Characteristics of planets studied.  $\tau_{orb}$ and $\tau_{rot}$ are the orbital and rotational periods, respectively.  $T_{n}$ is the equilibrium night-side temperature (see \S 2 and 3).  The other variables have their usual meanings.  The inclination $i$ is known only for the transiting planet HAT-P-2 b.}
\label{planettab}
\end{table}

\begin{table}
\begin{center}
\begin{tabular}{|r|c|c|c|c|c|}
\hline  &$M_*/M_\odot$&$R_*/R_\odot$&$T_* (K)$  	&$L_*/L_\odot$&$V$\\ 
\hline  HD 80606		&1.0		&1.05	&5573	&0.947	&9.06\\
\hline HD 37605 		&0.800	&0.967	&5391	&0.707	&8.67\\
\hline HD 108147 		&1.19	&1.29	&6156	&2.14	&6.99\\
\hline HAT-P-2			&1.35	&1.80	&6290	&4.25	&8.71\\
\hline HD 118203 		&1.23	&2.13	&5600	&4.00	&8.05\\
\hline  HD 185269		&1.28	&2.01	&5980	&4.61	&6.67\\
\hline
\end{tabular}
\end{center}
\caption{Characteristics of parent stars.  Only the radius $R_*$, the temperature $T_*$, and the luminosity $L_*$ are directly used in computing the light curves;  however, the mass $M_*$ and the apparent magnitude $V$ are also included for reference.}
\label{startab}
\end{table}

\begin{table}
\begin{center}
\begin{footnotesize}
\begin{tabular}{|r|c|c|c|c|c|c|c|}
\hline &$T_{min}$ (K) &$T_{max}$ (K)&$F_{min}/F_*$&$F_{max}/F_*$&$\Delta P / P$&$v_{max}$ \small{(m s$^{-1})$}&$v_{rms}$ \small{(m s$^{-1})$}\\
\hline  HD 80606 b&700&1200&$3.8\cdot10^{-4}$&$8.0\cdot10^{-4}$&0.2&670&200\\
\hline HD 37605 b&340&690&$7.1\cdot10^{-5}$&$2.6 \cdot 10^{-4}$&--&300&70\\
\hline HD 108147 b&680&1400&$1.7\cdot10^{-4}$&$4.0 \cdot 10^{-4}$&--&1060&370\\
\hline HAT-P-2 b&950&2170&$3.2\cdot10^{-4}$&$9.0\cdot10^{-4}$&0.5&1460&360\\
\hline HD 118203 b&920&1770&$2.1\cdot10^{-4}$&$4.5\cdot10^{-4}$&--&1140&390\\
\hline  HD 185269 b&900&1760&$1.8\cdot10^{-4}$&$2.9\cdot10^{-4}$&--&720&1710\\
\hline
\end{tabular}
\end{footnotesize}
\end{center}

\caption{Summary of simulation results.  $T_{min}$ and $T_{max}$ are the global minimum and maximum temperatures, respectively; $F_{min}/F_*$ and $F_{max}/F_*$ are the minimum and maximum relative planetary fluxes in the 8 micron band.  $\Delta P / P$ gives the intensity of the over-pressure front on planets where an acoustic disturbance is present.  The maximum and root-mean-square wind speeds are shown in the last two columns.  The rms wind speed for HD 80606 b varies significantly over the course of the simulation due to the high eccentricity of its orbit; the peak value of $v_{rms}$ is given for this planet.}
\label{resultstab}
\end{table}

\end{document}